# A combination of capillary assembly and dielectrophoresis for wafer scale integration of carbon nanotubes-based electrical and mechanical devices.

*Florent Seichepine[a,b,c,d*], Sven Salomon[b,d], Maéva Collet [b,d], Samuel Guillon[b,d], Liviu Nicu[b,d], Guilhem Larrieu[b,d], Emmanuel Flahaut [a,c,d], Christophe Vieu[b,d]* a Université de Toulouse; UPS, INP; Institut Carnot Cirimat; 118, route de Narbonne, F-31062 Toulouse, cedex 9, France

b Université de Toulouse, UPS, INSA, INP, ISAE, LAAS F-31059, France

c CNRS, Institut Carnot Cirimat, F-31062 Toulouse, France

d CNRS-LAAS, 7avenue du colonel Roche, F-31077 Toulouse, France

Abstract : The wafer scale integration of carbon nanotubes (CNT) remains a challenge for electronic and electromechanical applications. We propose a novel CNT integration process relying on the combination of controlled capillary assembly and buried electrode dielectrophoresis (DEP). This process enables to monitor the precise spatial localization of a high density of CNTs and their alignment in a pre-defined direction. Large arrays of independent and low resistivity (4.4 x $10^{-5}$ Ω.m) interconnections were achieved using this hybrid assembly with double-walled carbon nanotubes (DWNT). Finally, arrays of suspended individual CNT carpets have been realized and we demonstrate their potential use as functional nano-electromechanical systems (NEMS) by monitoring their resonance frequencies (ranging between 1.7 MHz to 10.5MHz) using a Fabry-Perot interferometer.

**Introduction**

Carbon nanotubes (CNT) have been studied for nearly two decades as potential new material for advanced nanoelectronic systems.[1] Metallic CNT can be used as large current density carriers[2] while semiconducting CNT are studied for the next generation of field effect transistor.[3] Other applications such as gas sensors, infra-red sensors or nano-electromechanical systems (NEMS) have also been demonstrated.[4-8] Nevertheless, large scale integration of these devices is still undergoing research and efficient processes have to be developed. Patterning precisely CNT catalysts followed by catalytic chemical vapor deposition (CCVD) processes have already been achieved to locally grow vertical CNT forests. The structural quality of the CNT can be controlled by engineering the catalyst and optimizing growth conditions.[9, 10] Large scale assembly of complex CNT structures, such as three-dimensional microelectromechanical devices has been reported using localized growth.[11]

Yet, to ensure compatibility with complementary metal-oxide-semiconductor (CMOS) fabrication technologies, back-end process temperatures cannot exceed 400–450 °C [12]. These temperatures are not compatible with the usual CNT growth conditions without affecting dramatically their structural quality. On the other hand, *ex-situ* CNT can be precisely separated according to their size, helicity or electric properties due to recent advances in CNT selection protocols,[13] but novel manipulation techniques are required to integrate *ex-situ* synthesized CNT at the wafer scale.[14] Moreover, due to their specific aspect ratio, CNT properties are highly anisotropic and their orientation should be controlled for most applications where the objective is to take benefit from the intrinsic physical properties of these 1D nano-objects.

Large scale assemblies of oriented CNT based on the use of capillary forces, mechanical actions or chemical binding have already been reported.[15-17] These techniques are mainly focused on single devices or macroscale integration while assembly processes like the one relying on the use of dielectrophoretic forces (DEP) allows both selective deposition and orientation of CNT. DEP forces are induced by non uniform electric fields and act on polarisable micro or nano-size objects dispersed in a liquid. When using appropriate solvent and experimental parameters, CNT can be attracted by the strong electric field gradients.[18] Depending on the electrode configuration, CNT are either individually connected at large scale, using floating electrode DEP, or high densities of CNT are connected between DEP electrodes, using conventional DEP.[19-21] It has also been shown that the chirality of the attracted CNT can be controlled by tuning the DEP parameters[22] and that these processes can be up-scaled for wafer scale implementation.[23, 24] Nevertheless, current DEP processes use the DEP electrodes both as assembling and actuating electrodes, which represents a severe limitation for their practical usability. Indeed, a complex interconnection design between several assembly sites or the independent use of a single device is impossible.

In this paper, we describe a versatile way to integrate dense layers of oriented CNT, at the wafer-scale, using buried electrode-dielectrophoresis coupled to capillary assembly. The CNT are oriented along electric field lines and assembled at pre-defined positions with adjustable density by controlling independently the capillary and DEP forces. During the whole process, CNT are electrically insulated from the DEP electrodes. CNT assembly sites are then connected individually using a dedicated metallization step which design is totally unconstrained. Using this original methodology, we demonstrate wafer-scale integration of oriented CNT into various kinds of elementary devices. By implementing a last etching process step, we demonstrate a NEMS proof of concept where the mechanically resonant structures are made of dense but thin carpets of oriented CNT. These devices are called carbon NEMS for simplification in the following.

**Results and discussion.**

**Process flow.** A schematic view of the process is presented on Figure 1. A silicon wafer is thermally oxidized (300nm) to avoid any conduction through the bulk substrate. Chromium (10nm) / Gold (90nm) DEP electrodes are patterned by photolithography using a conventional lift-off process. Chromium is used as an adhesion layer

DEP electrode shapes have been designed to create localized strong electric field gradients between the inter-electrode gaps. Interdigitated geometry was used to create an array of 40x18 gaps per square centimeter. Several electrode shapes have been tested: rectangular or triangular electrodes as seen in the inset on Figure 2. The electrode width ranged from 2 to 50 µm and the inter-electrode gap ranged from 2 to 12 µm.

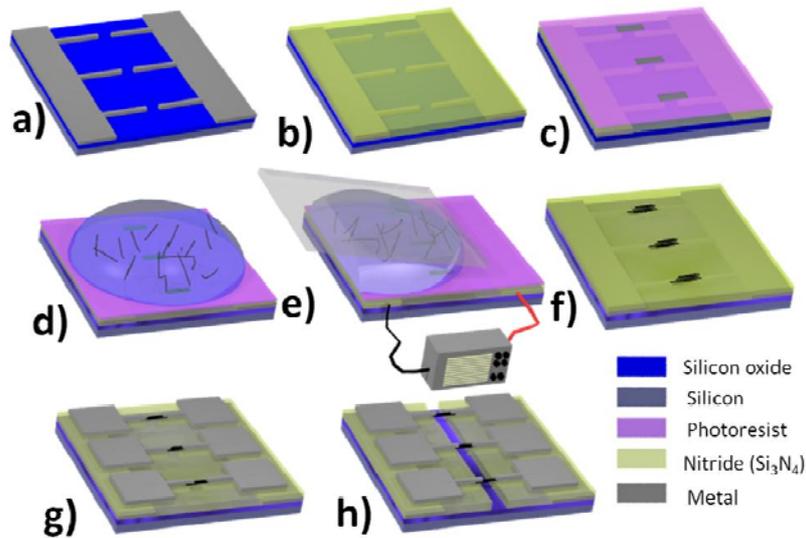

Figure 1 : Schematic view of the complete process used to integrate CNT dispersed in solution by a combination of DEP and capillary assembly. a) Cr (10nm)/Au(90nm) DEP electrodes are patterned on a thermally oxidized (300nm) silicon wafer. b) A 300 nm thick silicon nitride ($Si_3N_4$) layer is deposited by PECVD on the electrodes. c) A photoresist is spin-coated on the silicon nitride and cavities are patterned on the top of the DEP electrodes by proximity UV photolithography. d) A droplet of homogeneously dispersed CNT suspension is poured on the top of the photoresist. e) The function generator is used to deliver a localized electric field between the DEP electrodes while the meniscus is scanned at a controlled velocity over the patterns using a capillary assembly system. f) After assembly the photoresist is removed. g) The CNT connections are individually connected by Cr (10nm)/Au(90nm) independent electrodes pairs. h) The $Si_3N_4$ layer can be selectively etched by HF and critical point drying (CPD) for generating Carbon NEMS array.

A 300nm silicon nitride (Si$_3$N$_4$) layer was deposited by Plasma Enhanced Chemical Vapor Deposition (PECVD) on the DEP electrodes in order to separate the DEP electrodes from the CNT suspension (Figure 1 b). As the CNT suspension is electrically insulated from the DEP electrode, the CNT are not damaged during the assembly process and electrochemical reactions between the DEP electrodes are avoided even when applying high voltages. A photoresist layer (500nm AZ Lor 3A photoresist) was spin coated and patterned with cavities above the gaps of the DEP electrodes (Figure 1 c). The geometry of these cavities can be tuned to control the shape, length and width of the final CNT connections. A 100µl droplet of well dispersed CNT suspension was poured on the photoresist. The function generator was plugged to the DEP electrodes to deliver localized strong electric field gradients. The meniscus was scanned at a controlled velocity over the patterns and under electrical stimulation of the buried electrodes. (Figure 1 d and Figure 1 e). The photoresist layer was then properly removed. A lift-off process was used to pattern independent pairs of electrodes (Cr 10nm /Au 90nm). Those electrodes were designed to connect the CNT assembly sites individually (Figure 1 g). A SEM image of a typical wafer prepared using this process-flow is presented on Figure 2. Both individual electrode pairs (brighter) and buried DEP electrodes (bright grey) are visible. Finally, the CNT connections can be suspended by selective wet etching of the silicon nitride layer under the CNT carpet using hydrofluoric acid solution (5%, 1min30) followed by a critical point drying (CPD) step (Figure 1 h). This last step generates a large array of carbon NEMS.

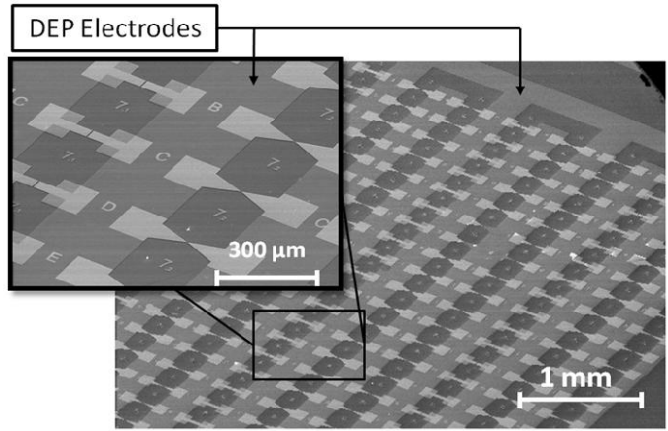

Figure 2: SEM image of a large array of individual electrodes connecting the CNT carpet (Figure 1 step g). Different gaps and shapes were used for the connection electrodes. The DEP electrodes can be seen under the $Si_3N_4$ layer in bright grey.

**Dielectrophoresis and capillary assembly.** CNT can be approximated by prolate spheroids[25] as their cylindrical shape exhibit a high aspect ratio. This approximation allows using the analytical expression of the DEP force [26]:

$$\vec{F}_{DEP} = \frac{2}{3} \pi \varepsilon_m l r^2 \operatorname{Re}[F(\omega)] \nabla \vec{E}^2$$

where $\varepsilon_m$ is the medium permittivity, $l$ is the CNT length, $r$ is the CNT radius, $\nabla \vec{E}$ is the gradient of the electric field and $F(\omega)$ is the Clausius Mossotti (CM) factor. The CM factor is determined by the frequency, the permittivities and conductivities of the medium and the CNT. In order to attract CNT towards high electric field gradients, the CM factor has to be positive which is achieved when using a 100 KHz signal. Therefore, the DEP assembly process was performed using a 40Vpp sine signal at 100 kHz. The highest electric field gradient is observed at the inter electrode gap. Yet, CNT parasitic presence at electrode edges has also been observed because of the long axis DEP force effect.[27]

When assembling the CNT using only DEP followed by solvent evaporation it was not possible to obtain a high density of assembled CNT. In fact, the DEP force is rapidly vanishing above the DEP electrode plane. Moreover, the very first landed CNT shield the electric field and lower the DEP force. To circumscribe this phenomenon, we have coupled the DEP effect with another methodology allowing the manipulation of the CNT suspension for concentrating the nanotubes precisely on the DEP gaps while keeping them trapped in a very thin liquid layer. For this purpose, capillary assembly appeared as a method of choice.[28-30] Indeed, many experimental and theoretical works dedicated to the directed assembly of colloid solution on patterned substrates have shown how to control the assembly process parameters.[31, 30] The fine tuning of temperature and meniscus displacement over the patterns can control the pinning of the triple line on the topographical patterns. This pinning leads to a local enrichment of the suspension on these areas due to convecting flows directed towards the thin layer of liquid stretched over the pining sites.[32, 33] The CNT manipulation and alignment by capillary assembly technique have been proved using topological[34] and temperature control.[35] In our case, the contact angle of 20° to 25° and assembly performed at 25°C with a scanning velocity of the meniscus of 10µm.s$^{-1}$ parallel to the cavities length turned out to concentrate the CNT inside the photoresist cavities just on the top of the DEP electrodes. When the capillary assembly was performed without DEP stimulation, large amounts of disorganized and not oriented CNT were observed in the cavities.

Due to the combined use of DEP and capillary assemblies, it was possible to generate organized deposition of CNT exhibiting good orientation and high density if needed. Finally, in this process, the alignment of the CNT can be controlled through the electrical parameters of the DEP set-up (bias, frequency) while the density of the trapped elements could be quasi-independently adjusted by tuning the parameters of the capillary assembly set-up (T°, scanning speed, contact angle, shape of the cavities…).

The assembly parameters were chosen to create dense carpets of oriented CNT. A suspension containing 10mg.L$^{-1}$ of double-walled carbon nanotubes (DWNT) and 10 mg.L$^{-1}$ of a dispersing agent (Carboxymethyl cellulose, CMC) was used. The speed of the capillary assembly was set to 10 µm.s$^{-1}$. The function generator supplied a 100kHz and 400mVpp sine signal which was amplified 100 times. Large scale characterizations were performed in order to measure the CNT orientation, the density between each connection and the structural quality of the integrated CNT. The selectivity of the deposition process has been characterized using SEM imaging (Figure 3) and AFM measurements (Figure 4) at different processing steps.

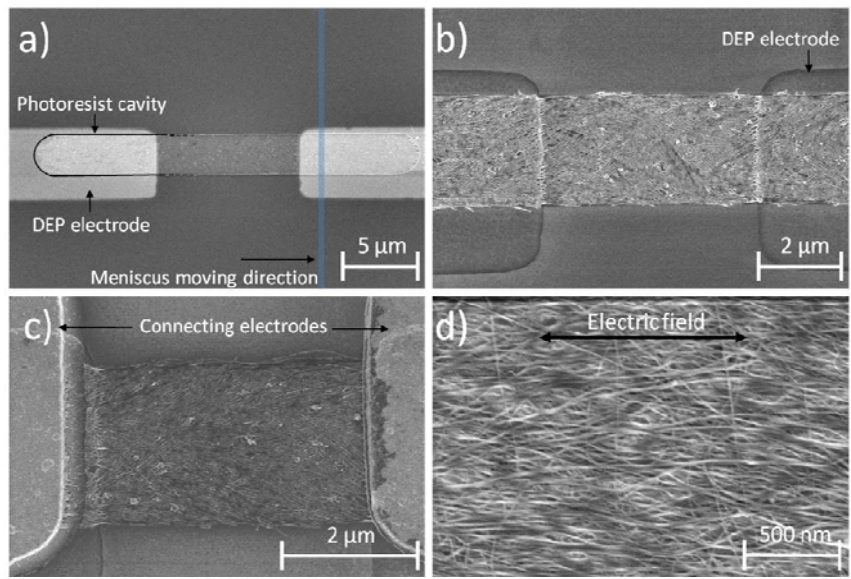

Figure 3 : SEM images of the DWNT connections at different processing steps. a) A 26µm-long cavity filled with a dense DWNT layer. The DEP electrodes are visible in bright with a gap of 10µm. b) DWNT layer on the Si$_3$N$_4$ film after removal of the resist (Figure 1 f). Buried DEP electrodes spaced by 5µm are visible. c) A DWNT layer with connecting gold electrodes (5µm gap) (step g)). d) Typical morphology of the obtained DWNT carpet.

A typical SEM image of the assembled CNT layer after DEP/Capillary assembly is shown in Figure 3 a, while the same device after careful photoresist removal is shown on Figure 3 b. AFM was used to measure the height of the CNT patterns. The height of the CNT layer turns out to depend on the size of the DEP electrode gap. An average height of about 10nm was measured for the 2µm gaps and decreased to 5 nm for the 10 µm gap. It has been previously shown[36] that the diameter of the used DWNT ranges from 1 to 3 nm. We can conclude that our CNT carpets are made of an average of 1 to 4 layers of CNT mainly packed into bundles. We explain this small number by the fact that the electric field is shielded by the first CNT which are attracted towards the $Si_3N_4$ layer. The individual electrodes were then patterned to connect each assembled CNT layer. The AFM and SEM images of such connections are shown in Figure 3 c and Figure 4, respectively.

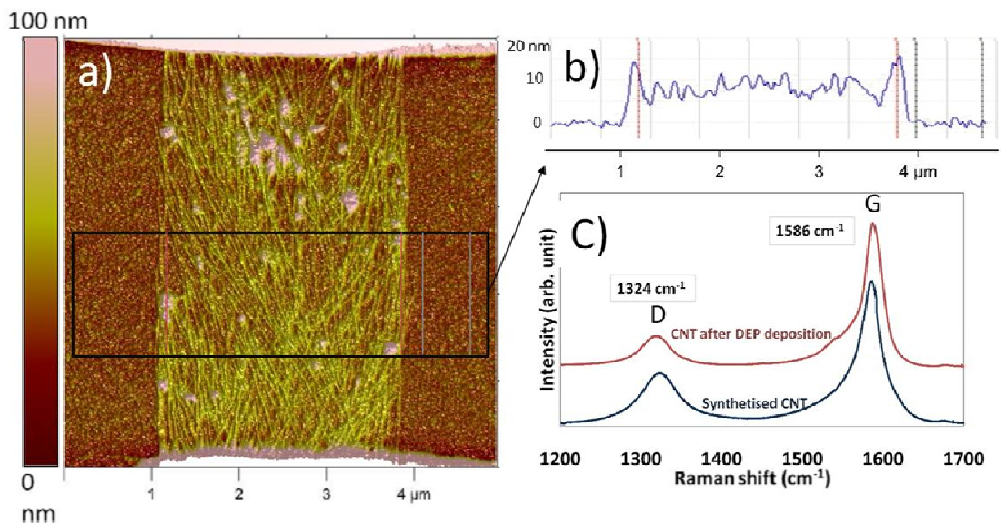

Figure 4: a) AFM image of a typical DWNT connection. The connecting electrodes are on the top and bottom of the image. b) Height profile of the DWNT carpet (black box in a)) (average of 9.2nm). c) Averaged Raman spectra (633nm) of the DWNT before (as-synthesized) and after the DEP deposition.

Raman spectroscopy was used to control the structural quality of the CNT through the measurement of the ratio between the intensity of the G band (IG) and the D band (ID) using a 633nm laser.[37] The Raman analysis of these connections shows very intense peak at 1324 cm$^{-1}$ and 1586 cm$^{-1}$ which are related to the D and G bands, respectively. The Figure 4 c) is an average of Raman spectra of a sample of CNT before and after completing the assembly process. The ID/IG ratio is 0.16. Very interestingly, we have observed that this ratio was systematically smaller when compared to the spectra obtained from the dried droplet of the suspension before integration. This result shows that a kind of selection is operated during the assembly, favoring the trapping of CNT of high structural quality. We explain this phenomenon by the specificity of DEP forces with respect to the electronic properties of CNT.

After patterning the connecting electrode level, large scale electrical characterizations were performed. Depending on the electrode shape and the deposition parameters, the yield of connected electrodes ranged from about 70 to 95%. The resistance was ranging from about 300 Ω for 50μm width and 2μm gap connecting electrodes to about 25 kΩ for the 4μm width 12μm gap connecting electrodes. The average resistivity of those CNT connections ranged from 4.4 x 10$^{-5}$ Ω.m for the denser to 5 x 10$^{-4}$ Ω.m for the 12μm long where the density is lower. This low resistivity compared to other CNT aligned layers described in the literature is in accordance with the high density and orientation score of our layers.[1, 7, 11]

For demonstrating the wafer scale integration of these carpets of high structural quality CNT into functional NEMS devices, we have optimized a release process through the wet etching of the silicon nitride layer followed by a CPD procedure. A tilted view of a typical device is shown on Figure 5. After fabrication, the mechanical oscillations of the structures were analyzed using a dedicated Fabry-Perot interferometry set-up which allows the measurement of the resonance frequency and corresponding quality factor under vacuum.[38] Depending on the length of the CNT carpet a resonance frequency from 1.7 MHz for the 10µm length to 10.5 MHz for the 5µm length was measured. Quality factors ranging from 10 to 400 have been measured for several devices. As expected, the resonance frequency is decreasing when the gap increases. However, the adsorbed impurities and variations of the CNT densities may induce a significant dispersion in the resonance frequency measurements.

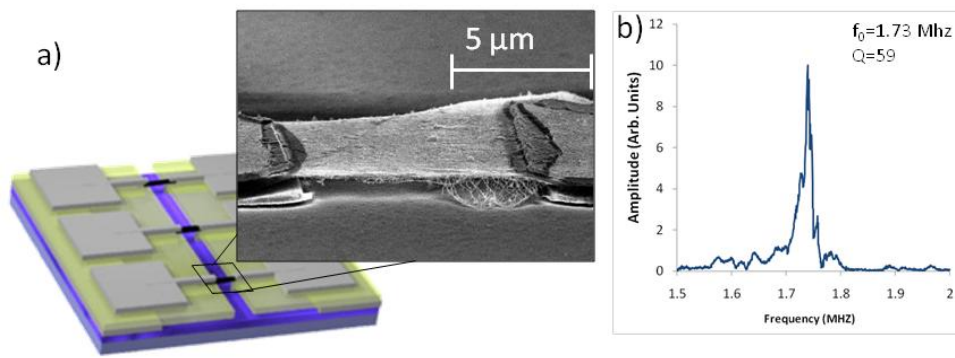

Figure 5: a) SEM image of a suspended CNT carpet. b) Typical Fabry-Perot measurement of one device (Length: 10µm, thickness: 3nm, width: 8µm).

In conclusion, we have developed an effective technique for large-scale integration of oriented carbon nanotubes inside functional devices. This technique based on capillary assembly and buried dielectrophoresis electrodes allows aligning and assembling with high accuracy dense CNT carpets. By using this protocol large arrays of electrically independent devices such as interconnections and CNT NEMS have been realized. SEM and AFM analysis showed thin, dense, aligned and precisely located CNT layers. A Fabry-Perot measurement set-up has been used to measure the resonance frequency of the suspended structures.

**Materials and methods**

The CNT used for this work were mainly double-walled carbon nanotubes (DWNT). They have been prepared by Catalytic Chemical Vapor Deposition (CCVD) by the reaction of decomposition of $CH_4$ at 1000°C ($H_2$:$CH_4$ atm.), as reported earlier.[39] The catalyst was then removed by non-oxidizing HCl treatment. Transmission electron microscopy proved that the sample contained approximately 80% of DWNTs, the rest being mainly SWNTs (~15%) and Triple-Walled Nanotubes.[36] The DWNT were properly dispersed using deionized water with Carboxymethyl cellulose (CMC).[40] To avoid any CNT agglomeration, the suspension was homogenized using ultrasonic tip and then centrifugation. The suspensions obtained were well dispersed and stable. No additional purification step was performed for this study. Therefore our sample still contained some disorganized carbon species (as-synthesized CNT).

A function generator (Tektronix AFG3102) was used to supply a controlled sine voltage. Signals are amplified with a FLC A800 amplifier (bandwidth from DC to 250 kHz). In order to control the capillary forces the substrate was placed on a motorized translation stage below a fixed glass slide at a distance of few hundreds of microns. The substrate was dragged slowly parallel to the inter electrode gap at a precisely controlled speed of 10 $\mu m.s^{-1}$.[41]

Raman spectroscopy was performed using a Labram HR 800 from Horiba Jobin Yvon with a 633nm (He/Ne) laser. SEM images were obtained on a S-4800 FEG-SEM from Hitachi with a 1kev to 30kev acceleration voltage depending on the sample. AFM measurements were realized with a Nanoscope 3000 from Veeco. The resistivity measurements have been performed on a Station Karl Suss PA200 and an Agilent 4142B Modular DC Source/Monitor.

The measurements of the nanotubes bridges resonance frequencies were achieved using a Fabry-Perot interferometry bench measurement originally developed for the characterization of silicon NEMS. The samples were mounted on a small piezoelectric pellet electrically actuated by a network analyzer coupled with a high frequency amplifier so that the entire substrate was vibrating. The device was placed inside a vacuum chamber pumped down to $2 \times 10^{-6}$ mbar at room temperature. A 30 mW He-Ne laser was focused on the structure through the chamber window using a beam expander and a long working distance microscope objective. As in a Fabry-Perrot cavity, the laser beam is partially reflected by the structure and partially by the substrate generating interferences which are detected by a photodetector connected to the network analyzer that allows closed loop measurement.**Erreur ! Source du renvoi introuvable.** The vibration of the bridge generates a periodic variation in the gap size between the bridge and the substrate which produces a periodic change in the interference intensity corresponding to the oscillation frequency of the nanostructures.

[1]     Close, G. F.; Yasuda, S.; Paul, B.; Fujita, S.; Wong, H.-S. P. A 1 GHz Integrated Circuit with Carbon Nanotube Interconnects and Silicon Transistors. *Nano Letters* **2008**, *8*, 706–709

[2]     Dai, H.; Wong, E. W.; Lieber, C. M. Probing Electrical Transport in Nanomaterials: Conductivity of Individual Carbon Nanotubes. *Science* **1996**, *272*, 523–526

[3]     Martel, R.; Schmidt, T.; Shea, H. R.; Hertel, T.; Avouris, P. Single- and multi-wall carbon nanotube field-effect transistors. *Applied Physics Letters* **1998**, *73*, 2447–2449

[4]     Lassagne, B.; Bachtold, A. Carbon nanotube electromechanical resonator for ultrasensitive mass/force sensing. *Comptes Rendus Physique* **2010**, *11*, 355 – 361

[5]     Arun, A.; Campidelli, S.; Filoramo, A.; Derycke, V.; Salet, P.; Ionescu, A. M.; Goffman, M. F. SWNT array resonant gate MOS transistor. *Nanotechnology* **2011**, *22*, 055204

[6]     Avouris, P.; Chen, Z.; Perebeinos, V. Carbon-based electronics. *Nat Nano* **2007**, *2*, 605–615

[7]     Acquaviva, D.; Arun, A.; Esconjauregui, S.; Bouvet, D.; Robertson, J.; Smajda, R.; Magrez, A.; Forro, L.; Ionescu, A. M. Capacitive nanoelectromechanical switch based on suspended carbon nanotube array. *Applied Physics Letters* **2010**, *97*, 233508

[8]     Sazonova, V.; Yaish, Y.; Ustunel, H.; Roundy, D.; Arias, T. A.; McEuen, P. L. A tunable carbon nanotube electromechanical oscillator. *Nature* **2004**, *431*, 284–287

[9] Cantoro, M.; Hofmann, S.; Pisana, S.; Scardaci, V.; Parvez, A.; Ducati, C.; Ferrari, A. C.; Blackburn, A. M.; Wang, K.-Y.; Robertson, J. Catalytic Chemical Vapor Deposition of Single-Wall Carbon Nanotubes at Low Temperatures. *Nano Letters* **2006**, *6*, 1107–1112

[10] Yan, F.; Zhang, C.; Cott, D.; Zhong, G.; Robertson, J. High-density growth of horizontally aligned carbon nanotubes for interconnects. *phys. stat. sol. (b)* **2010**, *247*, 2669–2672

[11] Hayamizu, Y.; Yamada, T.; Mizuno, K.; Davis, R. C.; Futaba, D. N.; Yumura, M.; Hata, K. Integrated three-dimensional microelectromechanical devices from processable carbon nanotube wafers. *Nat Nano* **2008**, *3*, 289–294

[12] Morgen, M.; Ryan, E. T.; Zhao, J.-H.; Hu, C.; Cho, T.; Ho, P. S. Low dielectric constant materials for VLSI interconnects. *Annual Review of Materials Science* **2000**, *30*, 645–680

[13] Arnold, M. S.; Green, A. A.; Hulvat, J. F.; Stupp, S. I.; Hersam, M. C. Sorting carbon nanotubes by electronic structure using density differentiation. *Nat Nano* **2006**, *1*, 60–65

[14] Wang, C.; Zhang, J.; Ryu, K.; Badmaev, A.; De Arco, L. G.; Zhou, C. Wafer-Scale Fabrication of Separated Carbon Nanotube Thin-Film Transistors for Display Applications. *Nano Letters* **2009**, *9*, 4285–4291

[15] Kang, S. J.; Kocabas, C.; Kim, H.-S.; Cao, Q.; Meitl, M. A.; Khang, D.-Y.; Rogers, J. A. Printed Multilayer Superstructures of Aligned Single-Walled Carbon Nanotubes for Electronic Applications. *Nano Letters* **2007**, *7*, 3343–3348

[16] Yu, G.; Cao, A.; Lieber, C. M. Large-area blown bubble films of aligned nanowires and carbon nanotubes. *Nat Nano* **2007**, *2*, 372–377

[17] Choi, S.-W.; Kang, W.-S.; Lee, J.-H.; Najeeb, C. K.; Chun, H.-S.; Kim, J.-H. Patterning of Hierarchically Aligned Single-Walled Carbon Nanotube Langmuirâˆ'Blodgett Films by Microcontact Printing. *Langmuir* **2010**, *26*, 15680–15685

[18] Duchamp, M.; Lee, K.; Dwir, B.; Seo, J. W.; Kapon, E.; Forro, L.; Magrez, A. Controlled Positioning of Carbon Nanotubes by Dielectrophoresis: Insights into the Solvent and Substrate Role. *ACS Nano* **2010**, *4*, 279–284

[19] Shekhar, S.; Stokes, P.; Khondaker, S. I. Ultrahigh Density Alignment of Carbon Nanotube Arrays by Dielectrophoresis. *ACS Nano* **2011**, *5 (3)*, 1739–1746

[20] Krupke, R.; Hennrich, F.; Kappes, M. M.; V. Löhneysen, H. Surface Conductance Induced Dielectrophoresis of Semiconducting Single-Walled Carbon Nanotubes. *Nano Letters* **2004**, *4*, 1395–1399

[21] Vijayaraghavan, A.; Blatt, S.; Weissenberger, D.; Oron-Carl, M.; Hennrich, F.; Gerthsen, D.; Hahn, H.; Krupke, R. Ultra-Large-Scale Directed Assembly of Single-Walled Carbon Nanotube Devices. *Nano Letters* **2007**, *7*, 1556–1560

[22] Krupke, R.; Hennrich, F.; Löhneysen, H. V.; Kappes, M. M. Separation of Metallic from Semiconducting Single-Walled Carbon Nanotubes. *Science* **2003**, *301*, 344–347

[23] Monica, A. H.; Papadakis, S. J.; Osiander, R.; Paranjape, M. Wafer-level assembly of carbon nanotube networks using dielectrophoresis. *Nanotechnology* **2008**, *19*, 085303

[24] Vijayaraghavan, A.; Hennrich, F.; Strüzl, N.; Engel, M.; Ganzhorn, M.; Oron-Carl, M.; Marquardt, C. W.; Dehm, S.; Lebedkin, S.; Kappes, M. M.; Krupke, R. Toward Single-Chirality Carbon Nanotube Device Arrays. *ACS Nano* **2010**, *4*, 2748–2754

[25] Venermo, J.; Sihvola, A. Dielectric polarizability of circular cylinder. *Journal of Electrostatics* **2005**, *63*, 101 – 117

[26] Jones, T. *Electromechanics of Particles*; Cambridge University press: New York, NY, 1995

[27] Benedict, L. X.; Louie, S. G.; Cohen, M. L. Static polarizabilities of single-wall carbon nanotubes. *Phys. Rev. B* **1995**, *52*, 8541–8549

[28] Bowden, N.; Terfort, A.; Carbeck, J.; Whitesides, G. M. Self-Assembly of Mesoscale Objects into Ordered Two-Dimensional Arrays. *Science* **1997**, *276*, 233–235

[29] Yin, Y.; Lu, Y.; Gates, B.; Xia, Y. Template-Assisted Self-Assembly : A Practical Route to Complex Aggregates of Monodispersed Colloids with Well-Defined Sizes, Shapes, and Structures. *Journal of the American Chemical Society* **2001**, *123*, 8718–8729

[30] Xia, Y.; Yin, Y.; Lu, Y.; McLellan, J. Template-Assisted Self-Assembly of Spherical Colloids into Complex and Controllable Structures. *Adv. Funct. Mater.* **2003**, *13*, 907–918

[31] Kralchevsky, P. A.; Denkov, N. D. Capillary forces and structuring in layers of colloid particles. *Current Opinion in Colloid & Interface Science* **2001**, *6*, 383–401

[32] Malaquin, L.; Kraus, T.; Schmid, H.; Delamarche, E.; Wolf, H. Controlled Particle Placement through Convective and Capillary Assembly. *Langmuir* **2007**, *23*, 11513–11521

[33] Cerf, A.; Molnar, G.; Vieu, C. Novel Approach for the Assembly of Highly Efficient SERS Substrates. *ACS Applied Materials & Interfaces* **2009**, *1*, 2544–2550

[34] Xiong, X.; Jaberansari, L.; Hahm, M. G.; Busnaina, A.; Jung, Y. J. Building Highly Organized Single-Walled-Carbon-Nanotube Networks Using Template-Guided Fluidic Assembly. *Small* **2007**, *3*, 2006–2010

[35] Ye, Z.; Lee, D.; Campbell, S. A.; Cui, T. Thermally enhanced single-walled carbon nanotube microfluidic alignment. *Microelectronic Engineering* **2011**, *88*, 2919–2923

[36] Flahaut, E.; Bacsa, R.; Peigney, A.; Laurent, C. Gram-scale CCVD synthesis of double-walled carbon nanotubes. *Chemical Communications* **2003**, 1442–1443

[37] Osswald, S.; Flahaut, E.; Ye, H.; Gogotsi, Y. Elimination of D-band in Raman spectra of double-wall carbon nanotubes by oxidation. *Chemical Physics Letters* **2005**, *402*, 422 – 427

[38] Carr, D. W.; Craighead, H. G. Fabrication of nanoelectromechanical systems in single crystal silicon using silicon on insulator substrates and electron beam lithography. *J. Vac. Sci. Technol. B* **1997**, *15*, 2760–2763

[39]  Flahaut, E.; Laurent, C.; Peigney, A. Catalytic CVD synthesis of double and triple-walled carbon nanotubes by the control of the catalyst preparation. *Carbon* **2005**, *43*, 375 – 383

[40]  Minami, N.; Kim, Y.; Miyashita, K.; Kazaoui, S.; Nalini, B. Cellulose derivatives as excellent dispersants for single-wall carbon nanotubes as demonstrated by absorption and photoluminescence spectroscopy. *Applied Physics Letters* **2006**, *88*, 093123

[41]  Cerf, A.; Thibault, C.; Geneviève, M.; Vieu, C. Ordered arrays of single DNA molecules by a combination of capillary assembly, molecular combing and soft-lithography. *Microelectronic Engineering* **2009**, *86*, 1419 – 1423